\documentclass[12pt]{article}
\usepackage{times}
\usepackage{geometry}
\usepackage[utf8]{inputenc}
\usepackage{enumitem,amssymb}
\usepackage{ragged2e}
\newlist{thematic}{itemize}{8}
\setlist[thematic]{label=$\square$}
\usepackage{pifont}
\newcommand{\cmark}{\ding{51}}%
\newcommand{\done}{\rlap{$\square$}{\raisebox{2pt}{\large\hspace{1pt}\cmark}}%
\hspace{-2.5pt}}

\usepackage{fancyhdr}
\usepackage[USenglish]{babel}
\usepackage[utf8]{inputenc}
\usepackage[T1]{fontenc}
\usepackage{epsfig}
\usepackage{wrapfig}
\usepackage{color}

\usepackage[vflt]{floatflt}
\usepackage{longtable}
\usepackage{caption}
\usepackage{jheppub}   
\usepackage[dvipsnames,svgnames,x11names]{xcolor}
\usepackage[all]{hypcap} 
\usepackage{amssymb,amsmath}
\usepackage{longtable}
\usepackage{amssymb}
\usepackage{amsmath, xspace}
\usepackage{graphics,graphicx} 
\usepackage{rotating}
\usepackage{color}
\usepackage{multirow}
\usepackage{subfigure}
\usepackage{sectsty}
\usepackage[outercaption]{sidecap}
\sidecaptionvpos{figure}{c}

\definecolor{DarkGreen}{rgb}{0.0, 0.3, 0.0}
\definecolor{purple}{rgb}{0.5, 0.0, 0.5}
\definecolor{red}{rgb}{1, 0.0, 0.0}
\definecolor{green}{rgb}{0, 1.0, 0.0}

\newcommand{\Moon}{$M_{Moon}$}                          

\newcommand{\apjl}{\rm ApJL}
\newcommand{\apjs}{\rm ApJS}

\newcommand{\aap}{\rm A$\&$A}

\def\ao{{\it Appl.~Opt.}}           

\def\3he{$^3{\rm He}$}
\def\arcdeg{\hbox{$^\circ$}}

\def\lsim{\mathrel{\lower2.5pt\vbox{\lineskip=0pt\baselineskip=0pt
           \hbox{$<$}\hbox{$\sim$}}}}

\def\gsim{\mathrel{\lower2.5pt\vbox{\lineskip=0pt\baselineskip=0pt
           \hbox{$>$}\hbox{$\sim$}}}}

\begin{document}
\raggedright
\huge
Astro2020 Science White Paper \linebreak
\smallskip
The hidden circumgalactic medium \linebreak
\normalsize

\noindent \textbf{Thematic Areas:} \hspace*{60pt} $\square$ Planetary Systems \hspace*{10pt} $\square$ Star and Planet Formation \hspace*{20pt}\linebreak
$\square$ Formation and Evolution of Compact Objects \hspace*{31pt} $\square$ Cosmology and Fundamental Physics \linebreak
  $\square$  Stars and Stellar Evolution \hspace*{1pt} $\square$ Resolved Stellar Populations and their Environments \hspace*{40pt} \linebreak
  $\done$    Galaxy Evolution   \hspace*{45pt} $\square$             Multi-Messenger Astronomy and Astrophysics \hspace*{65pt} \linebreak
  
\textbf{Principal Author:}

Name: Claudia Cicone
 \linebreak						
Institution:  INAF - Osservatorio astronomico di Brera, Milano (Italy)
 \linebreak
Email: claudia.cicone@inaf.it
 \linebreak
Phone: +39 02 72320381 
\smallskip

\textbf{Co-authors:} 
Carlos De Breuck (ESO); Chian-Chou Chen (ESO); Eelco van Kampen (ESO); Desika Narayanan (U.Florida); Tony Mroczkowski (ESO); Paola Andreani (ESO); Pamela Klaassen (UK ATC); Axel Weiss (MPIfR); Kotaro Kohno (U.Tokyo); Jens Kauffmann (MIT); Jeff Wagg (SKAO); Dominik Riechers (Cornell/MPIA); Bitten Gullberg (Durham); James Geach (U.Hertfordshire); Sijing Shen (U.Oslo); J. Colin Hill (IAS/CCA); Simcha Brownson (U.Cambridge).

 \captionsetup{labelformat=empty}
\begin{figure}[h]
   \centering
   \includegraphics[height=.7\textwidth,angle=90]{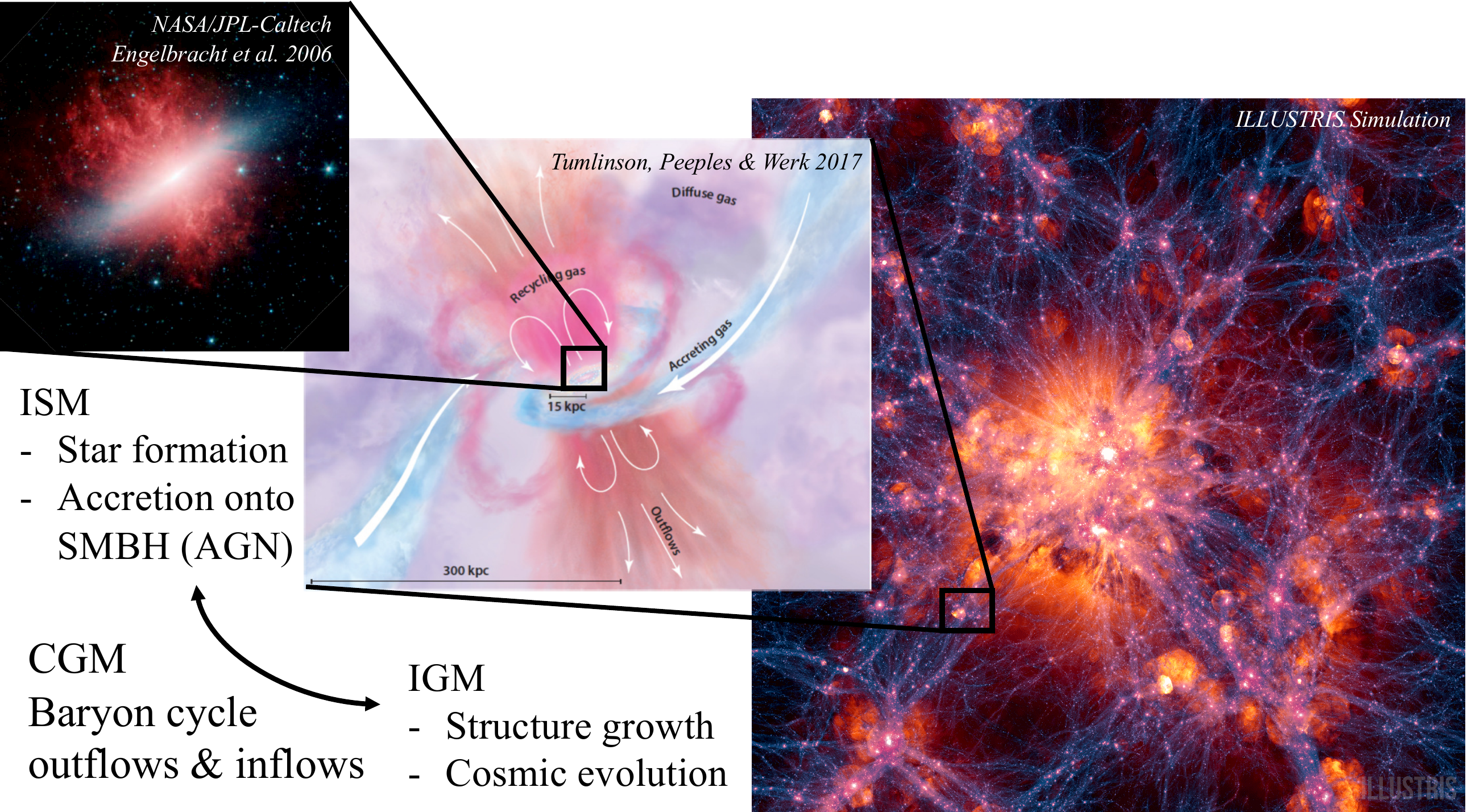}
   \caption{}
\end{figure}
\vspace{-15mm}

\setcounter{figure}{0}
\captionsetup{labelformat=default}

\justify
\textbf{Abstract:} 
The cycling of baryons in and out of galaxies is what ultimately drives galaxy formation and evolution. The circumgalactic medium (CGM) represents the interface between the interstellar medium and the cosmic web, hence its properties are directly shaped by the baryon cycle. Although traditionally the CGM is thought to consist of warm and hot gas, recent breakthroughs are presenting a new scenario according to which an important fraction of its mass may reside in the cold atomic and molecular phase. This would represent fuel that is readily available for star formation, with crucial implications for feeding and feedback processes in galaxies. However, such cold CGM, especially in local galaxies where its projected size on sky is expected to be of several arcminutes, cannot be imaged by ALMA due to interferometric spatial scale filtering of large-scale structures. We show that the only way to probe the multiphase CGM including its coldest component is through a large (e.g. 50-m) single dish (sub-)mm telescope.
\pagebreak
\section{Scientific context and motivation}

Galaxies grow and evolve within a network of dark matter through a finely-tuned exchange of baryons with the cosmic web, a process that is challenging to understand and model. The importance of describing baryons, their interplay and behaviour in and out of galaxies is two-fold. On the one hand, if our research were restricted to the cosmological framework, we would not be able to comprehend how the bright components of galaxies - the stars - formed out of dense gas clouds, hence how planetary systems originated, and eventually how life began. On the other hand, since the only observables against which we can test theoretical predictions are produced by baryons, a proper understanding of the complex mechanisms involving baryonic matter is essential to interpret the observable data and so test our cosmological paradigm. Hence, understanding the baryon cycle is an endeavour at the forefront of modern astrophysics, and it is relevant to many different fields, from Galactic studies to Cosmology. 

The baryon cycle problem can be visualised as follows: galaxies are  embedded in large gas haloes that we define ``circumgalactic medium (CGM)'', extending by hundreds of kpc (up to the virial radius). A small portion of CGM material, the interstellar medium (ISM), has already cooled and flowed towards the central regions of galaxies, and it is directly involved in the processes of star formation and accretion onto supermassive black holes (SMBHs). However, most of the CGM mass lies outside the ISM realm and represents the interface between the ISM and the matter distributed outside galaxies, the intergalactic medium (IGM). The CGM is fed - from the inside - by galactic outflows and fountains \citep{Biernacki+Teyssier18,Costa+18} - and, from the outside - by cosmic streams \citep{Keres+05,Dekel+09} and galaxy mergers \citep{Hani+18}. Furthermore, the CGM gets depleted via accretion onto galaxies, and through expulsion into the IGM by energetic feedback processes \citep{Silk+Rees98,DiMatteo+05,Hopkins+14}. The CGM is therefore intrinsically linked to the feedback and accretion processes in galaxies, and carries the imprint of their evolution. 

Although traditionally the CGM is thought to be dominated by hot and warm gas \citep{Tumlinson+17}, with little or no cold gas beyond the ISM scales, this picture has been challenged by recent breakthroughs in this field. {\it On the one hand, the lack of cold gas in most simulated galaxy haloes appears to be a bias of hydrodynamical simulations rather than a theoretical prediction}. More specifically, several studies are now finding that increasing the spatial resolution on halo scales (i.e. beyond the star forming ISM) naturally leads to an enhanced cold gas fraction in the CGM, because it allows to properly capture cooling processes and to spatially resolve smaller cold clumps embedded in a warmer medium \citep{Suresh+19,Hummels+18,vandeVoort+19}. {\it On the other hand, a massive cold ($T<10^4$~K) CGM reservoir may result from cosmic feeding, which at high redshift is dominated by cold filamentary inflows, but also from powerful feedback}. Indeed, recent studies have shown that intense episodes of star formation and active galactic nuclei (AGN) can drive massive ($\dot{M_{out}}\sim1000~M_{\odot}~yr^{-1}$) outflows of H{\sc i} and H$_2$ gas. These can extend by several kpc \citep{Cicone+14,Cicone+18b} and embed very dense ($n>10^4$~cm$^{-3}$) clouds \citep{Aalto+12,Aalto+15}. However, although H$_2$ outflows can be powerful enough to clear the central regions of galaxies within a few Myr, they are likely incapable of expelling all of the molecular ISM from the halo, both in starbursts \citep{Leroy+15} and in luminous quasars \citep{Fluetsch+19,Alatalo15}. Instead, cumulative effects of such outflows may be: (i) the transportation of material to CGM scales, where most of the (metal-enriched) H$_2$ and H{\sc i} gas stalls without escaping the halo \citep{Costa+18,Shen+13}, and (ii) the enhancement of gas cooling on halo scales \citep{Biernacki+Teyssier18,Prieto+17,Costa+15}. Therefore, we expect the CGM to be multiphase and entrain not only diffuse ionised gas, but also colder and denser clumps. Such cold and non-pristine gas represents fuel that, once (re)-accreted onto the ISM, would be readily available to feed star formation, and hence it would have a crucial role for galaxy evolution.

{\it As we argue in this paper, only (sub-)millimeter ((sub-)mm) and far-infrared (FIR) observations can probe the multiphase CGM including its coldest component, and so deliver a complete census of baryons in galaxy haloes. Such measurements are however extremely challenging even for ALMA because they require high sensitivity to large-scale emission and wide field of views (FoVs) which only a large (e.g. 50m) single-dish can provide.}  

\vspace{-3mm}
\section{State-of-the-art and forthcoming experiments}

Currently, we have very little observational constraints on the mass and physical properties of cold gas on CGM scales. The CGM of high-z (radio quiet) AGNs has so far been investigated predominantly in its diffuse ionised and atomic phases through absorption features detected in rest-frame UV spectra. Interestingly, some of these works hinted at a higher cold gas fraction in the CGM of luminous quasars at $z\sim1-3$ compared to inactive galaxies at similar redshifts and with similar masses, especially at large projected velocities of $|v|>300$~km~s$^{-1}$ \citep{Johnson+15}. Recently, the ubiquitous detection of low surface brightness $\gtrsim 50-100$~kpc-size Ly$\alpha$ nebulae, permitted by instruments such as MUSE at the VLT and KCWI at the Keck telescope, has enabled direct imaging of the CGM at $3<z<6$ \citep{Arrigoni-Battaia+19,Wisotzki+18,Borisova+16}. The properties of Ly$\alpha$ haloes around $z\geq3$ quasars suggest entrainment of cold ($T\lesssim10^{4}$~K) and dense (density, $n\gg1$~cm$^{-3}$) clumps up to $r\sim50$~kpc in these sources \citep{Arrigoni-Battaia+15}. However, the absorption of Ly$\alpha$ photons and the resonant nature of Ly$\alpha$ emission prevent a detailed kinematical study of the CGM using this tracer. Furthermore, the presence of a ionising source (such as a quasar) can significantly boost the Ly$\alpha$ flux, compared to what is expected, due to the ubiquitous fluorescent signal stimulated by the high UV background \citep{Cantalupo+12}. Therefore, Ly$\alpha$ imaging cannot provide an unbiased census of the high-z CGM. 

The ionised and neutral CGM, including any cold and dense H$_2$ gas, can be directly imaged through FIR fine structure lines (such as [C{\sc ii}]~158$\mu$m, [C{\sc i}]~609$\mu$m \& 370$\mu$m,  [N{\sc ii}]~205$\mu$m \& 122$\mu$m, [O{\sc iii}]~88$\mu$m, etc) and molecular rotational transitions (e.g. CO, HCN, CN, CS). At high-$z$, FIR lines are redshifted into the (sub-)mm atmospheric windows, allowing ground-based observations. Thanks to the bright and multiphase (H{\sc ii}, H{\sc i}, H$_2$) nature of [C{\sc ii}] \citep{Pineda+13}, sensitive [C{\sc ii}] line observations hold the potential to directly image faint and diffuse CGM components at $T<10^4$~K. Cicone et al. \citep{Cicone+15} demonstrated the existence of copious amounts of [C{\sc ii}]-emitting gas in the CGM of a luminous quasar at $z=6.4$, which was also found to host the most extended ($r\sim 30$~kpc)  high-velocity ($v\sim1400$~km~s$^{-1}$) outflow of cold gas known. By modelling the {\it uv} visibilities, 70\% of the [C{\sc ii}] flux at low projected velocities was found to trace a 20~kpc-scale nebula, while only 30\% of [C{\sc ii}] arises from the compact ISM. This result completely overturned the scenario favoured by earlier interferometric [C{\sc ii}] observations of the same source, which, lacking short antenna spacings, resolved out all of the extended emission. 

The cold CGM may be dense and bright enough to be detected also in CO line emission. The so-called `Spiderweb' galaxy, a protocluster at $z\sim2.2$ with a central bright radio galaxy, hosts a 70 kpc-size halo of molecular gas \citep{Emonts+16,Emonts+18}, which does not arise from individual galaxies but from the extended CGM. Surprisingly, its carbon abundance and excitation properties are similar to a normal star forming disk, hence supporting a `recycled' origin of the CGM - due to a combination of outflows, mass transfer among galaxies, gas accretion, and mergers. Interestingly, the CO(1-0) emission seen in the Spiderweb galaxy is so diffuse that it was resolved out by the JVLA, while being detected at high significance by the shorter baselines of ATCA \citep[see also][]{Emonts+18_ngVLA}.
Molecular haloes may not be unique to powerful quasars. Ginolfi et al. \citep{Ginolfi+17} discovered a 40 kpc-size CO(4-3) structure around a massive galaxy at $z=3.5$, where most of the molecular line emission is not associated with the ISM of the main galaxy or its satellites. {\it These pilot studies suggest that FIR and (sub-)mm line observations of the high-z CGM carried out with a high sensitivity to large-scale structures hold a huge discovery potential.}

The variety of (sub-)mm and FIR lines would allow us not only to detect any hidden massive CGM reservoir, but also to constrain its physical properties such as density, temperature, ionised-to-neutral gas fraction, ionisation parameter, and metallicity \citep{Fernandez-Ontiveros+16,Bethermin+16}. Moreover, the unique chemical properties of some molecular ions (OH$^+$, CH$^+$, H$_2$O$^+$) make them suitable to explore physical processes that can critically affect galaxy growth, such as turbulence and shocks \citep{Rangwala+14,Gonzalez-Alfonso+18}. These mechanisms are strictly associated to the baryon cycle (feedback through outflows and feeding by inflows) and so are particularly relevant to CGM gas. For example, ALMA imaging of CH$^+$ lines, detectable in absorption from diffuse molecular gas \citep{Godard+12}, has recently revealed hidden massive reservoirs of highly turbulent cool ($T\sim100$~K) gas, extending to $>10$~kpc, far beyond the $\sim1$~kpc-size ISM of (lensed) dusty star forming galaxies at $z\gtrsim2$ \citep{Falgarone+17}. The same observations detected extremely broad (FWZI$\sim2500~\rm km~s^{-1}$) CH$^+$ emission tracing very dense gas ($\rm n\sim10^5~cm^{-3}$) entrained in powerful galactic outflows, which are probably feeding the large-scale diluted turbulent H$_2$ CGM reservoir seen in absorption. 

In parallel, high sensitivity, wide FoV, broadband continuum observations will open interesting new avenues to the investigation of the hidden CGM. Bright high-z FIR/(sub-)mm continuum emitters can be used as background sources to detect diffuse, low column density CGM gas in absorption through virtually any molecular tracer (e.g. CO, HCN, HCO$^+$, OH, OH$^+$, CH$^+$, \citep{Muller+06,Combes08}). Such experiments require a large FoV to sample several continuum emitters and so overcome the likely low filling factor of the CGM absorbers. They also require high spectral resolution, since molecular line absorbers usually have line widths of $\sim10$~km~s$^{-1}$. 

Finally, sensitive continuum observations will enable the detection of the thermal Sunyaev-Zeldovich (tSZ) effect from the warm/hot phase of the CGM. The tSZ signal, being proportional to the integral of electron pressure, provides calorimetry of the ionised gas \citep[e.g.][]{Mroczkowski2019,Mroczkowski19b}. This, in turn, provides strong constraints on galaxy formation models, whose predictions span a wide range of tSZ signals depending on the AGN feedback prescriptions (see e.g.\ \citep{Battaglia2017}, and white paper by Battaglia \& Hill et al.).  
In addition, the integrated tSZ signal can give an estimate of the {\it total} mass of a system. Stacking of Planck, ACT, and SPT data \citep{Greco2015,Ruan2015} has allowed to reach down to galaxy mass scales, with typical fluxes of a few tens of $\mu$Jy\footnote{For example, a total mass of $M\sim10^{11}~M_{\odot}$ produces a negative continuum signal of $\sim -70~\mu$Jy at $z=0.15$.}.  Thus tSZ measurements of the CGM require good sensitivity and likely next generation (sub-)mm instruments.

\vspace{-3mm}
\section{Opportunities to make a big step forward}

{\bf Improving over ALMA: the quest for a large (sub)-millimetre single dish:}\newline
{\it The cold neutral CGM, and in particular the molecular CGM, is an almost completely unexplored avenue, due to the lack of (sub-)mm facilities that are able to detect and trace large-scale, low surface brightness structures.} The latter are indeed filtered out by interferometers, as explained below, and undetected by current single dish facilities because of their low sensitivity (e.g.\ the 12-meter APEX telescope or the ALMA Total Power antennas). Larger dishes such as the LMT~50-m or the IRAM~30-m telescopes have a limited frequency coverage, imposed by the atmospheric conditions at the telescope site as well as by the antenna surface accuracy which make observations at $\nu>345$~GHz~almost prohibitive.

ALMA has transformed our understanding of galaxies at both high and low redshift, but there are limitations to its capabilities. Due to spatial scale filtering, an interferometer only detects a fraction of the total flux density for sources with emission on scales larger than its shortest antenna spacing (baseline), even in mosaicked observations. The maximum recoverable scale (MRS) of an interferometer is $\sim0.6 \lambda/b_{min}$, where $b_{min}$ is the shortest baseline (close to the dish diameter), although the exact MRS is best determined through simulations because it depends on both the array configuration and the details of the large-scale structure. For the ALMA 12-m array, $\rm MRS\lesssim 3^{\prime\prime}$ at 850~GHz ($\rm MRS\lesssim8^{\prime\prime}$ for the ACA 7-m array) \citep[][]{ALMATec}. An MRS of $3^{\prime\prime}$ corresponds to a largest detectable angular scale of $\sim45$~pc at the distance of M~82 ($z=0.00073$, distance of 3~Mpc), and of $\sim0.6-3$~kpc at $z=0.01-0.05$, hence preventing the study of not only the CGM, but also of any large-scale ISM structure (such as a galactic outflow or a tidal tail) in nearby galaxies.  
However, at higher-$z$, a $\rm MRS\sim3^{\prime\prime}$ corresponds to
$\sim26-16$~kpc at $z=2-7$, which in principle allows some CGM science with ALMA (see $\S$~2) but only limited to much smaller structures compared to those detected in Ly$\alpha$ emission (with sizes of $\sim100$~kpc, e.g. \citep{Arrigoni-Battaia+19}). 


To further illustrate the advantage of a large single dish (SD) antenna vs ALMA in observing extended sources, we show here a sensitivity comparison. More specifically, we consider the ALMA 50 element array of 12-m diameter antennas, and compare it with a 50-m diameter SD antenna hypothetically placed near the ALMA site. The rms noise for an interferometric observation scales as $\sigma^{\rm interf}_{\rm rms}\sim T_{\rm sys}/[A_{\rm eff}(N_{\rm base}~\Delta \nu~t)^{1/2})]$, where $T_{\rm sys}$ is the noise contribution of the atmosphere and receiver, $A_{\rm eff}$ is the effective collecting area of each element, $t$ is the integration time, $\Delta \nu$ is the spectral bandwidth, and $N_{\rm base}$ is the number of baselines, which for $M$ array elements equals to $N_{\rm base} = M~(M-1)/2$ (for example, for the ALMA 12-m array and $M=50$ antennas, $N_{\rm base}=1225$). For a SD telescope, equipped with a heterodyne focal plane array (an instrument that delivers spectro-imaging information, i.e.\ 3D datacubes), the relation is similar, but we redefine $N$ to be the number of independent beams on the sky, so that $\sigma^{SD}_{\rm rms}\sim T_{\rm sys}/[A_{\rm eff}(N_{\rm beams}~\Delta \nu~t)^{1/2})]$. The effective area $A_{\rm eff}$ of an antenna scales with its geometric area, hence $\rm A_{eff}^{SD}=17.36~A_{eff}^{ALMA}$, and so $\sigma^{SD}$ matches $\sigma^{\rm ALMA}$ for $N_{\rm beams}=4$, if one assumes the same $\Delta \nu$,  $T_{\rm sys}$, and $t$. However, one should consider that the FoV of a 50m-antenna is smaller by a factor of $(50/12)^2$ than the FoV of ALMA. Hence, the sensitivities per FoV of ALMA and of a 50-m SD become comparable when the source area fills at least $\sim70$ beams. 
At 850 GHz (resolution of 1.45$^{\prime\prime}$) and for typical beam spacings (i.e.\ 2f$\lambda$, f=focal ratio), this is true for sources $>24^{\prime\prime}$ across, hence at larger sizes the mapping speed of the 50-m SD outperforms the full ALMA array.
Furthermore, we note that a SD telescope is much easier to keep at the forefront of upgraded technologies compared to a 50-element array. For instance, the new generation heterodyne receivers mounted on the APEX telescope already provide instantaneous spectral bandwidths (per polarization and per pixel) that are twice as large as the current ALMA bandwidth.

\begin{figure}[h!]
  \hspace{-4mm}
  \centering 
  \includegraphics[angle=90,scale=0.58]{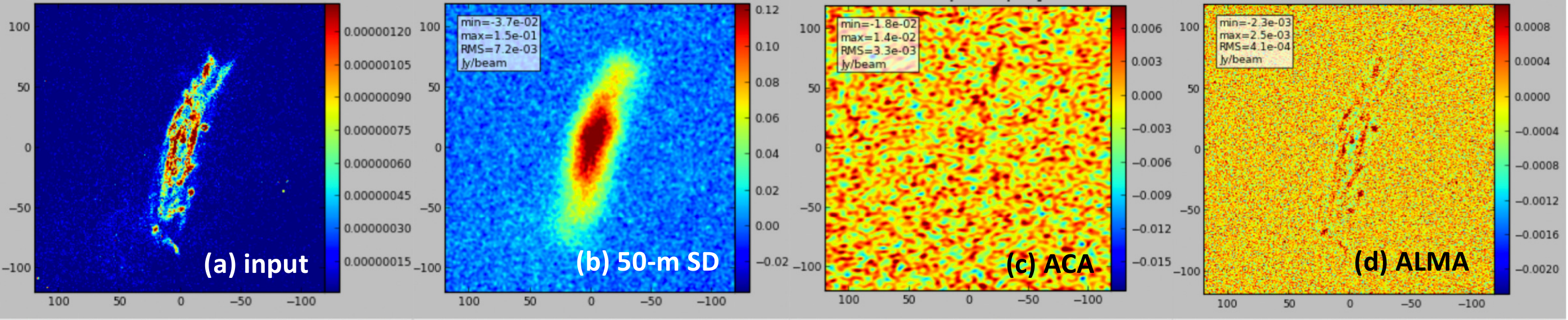}
   \caption{The expected molecular CGM of a star forming galaxy at $z=0.02$ as seen by  different (sub-)mm facilities in 10 hours of integration. The observing frequency corresponds to the CO(3-2) line (Band~7). Panel (a) shows the input image (details in main text). Panels (b-d) show CASA simulations corresponding to: (b) an on-the-fly map obtained with a 50-m SD telescope equipped with a single ALMA detector; (c) a mosaic map obtained with the ACA 7-m array; (d) a 576 pointings-mosaic map obtained with the ALMA 12-m array in compact configuration. 
   }\label{fig:sim}
\end{figure}

Figure~\ref{fig:sim} compares the capability of such hypothetical 50-m SD telescope (see, e.g. \cite{Bertoldi2018,deBreuck2018,Hargrave2018,Klaassen2018,Mroczkowski2018}) with that of the ACA and ALMA arrays in mapping large-scale emission from the cold CGM at $z=0.02$. The input image is drawn from the state-of-the-art SIMBA cosmological simulations with a $\rm (25/h  Mpc)^3$ volume \citep[][]{Dave+19}.  We selected a star-forming ($\rm SFR=11~M_{\odot}~yr^{-1}$) disk galaxy with stellar mass $M_*=1.3\sim10^{11}~M_{\odot}$, and total CGM gas mass $\rm 7.75\times10^{10}~M_{\odot}$. The equilibrium molecular line luminosities are calculated simultaneously by employing the thermal-chemical-radiative equilibrium code \texttt{DESPOTIC} \citep[][]{Krumholz14}. We then projected the particle information onto a uniform 1024$^3$ grid, and calculated the line-of-sight velocity-integrated intensities. The latter have been re-normalised to 1~Jy~km~s$^{-1}$, since our purpose is to compare how much flux density is recovered by the different facilities in $t=10$~hours. As shown in Fig.~\ref{fig:sim}, the molecular line flux is spread across a large projected area of $\sim200^{\prime\prime}$, corresponding to 80~kpc at $z=0.02$, which is not recovered by the interferometers. Clearly, for such low-z galaxies, ALMA and ACA are not optimised to probe the CGM because of (i) their small beam (even in the most compact configuration) which requires large mosaic observations to probe the full source extent (implying a lower integration time per pointing), and of the (ii) spatial scale filtering of large-scale emission, which affects even mosaic observations. The best results are obtained with the 50-m SD antenna, which we have conservatively assumed to be equipped with a single-pixel ALMA detector (but technology advancements will soon deliver much more sensitive multi-pixel detectors hence significantly shortening the SD observing time, \citep{Mroczkowski2018}).  

Of course, the most important factor to consider for ALMA vs a 50-m SD in the Atacama desert is their {\it complementarity}: on the one hand, as shown by Fig.~\ref{fig:sim}, ALMA will never access scales large enough to constrain the full extent - and mass - of the CGM - for this, single dish mapping is required. On the other hand, a 50-m SD will be limited to $1.5^{\prime\prime}$ spatial resolution at 850~GHz. Only joint SD {\it and} ALMA mapping can thus provide a complete view of the CGM over a broad range of spatial scales and cosmic times.

\vspace{-2mm}
\vspace{-2mm}


{\noindent \bf Synergies with astronomical facilities planned beyond 2030s:}\newline
Capabilities to map low surface brightness H{\sc i}~21cm emission from galaxies up to high-z, realised by (near-)future projects including the SKA \citep{SKA} and the ngVLA \citep{ngVLA}, will be highly complementary to the (sub-)mm-based studies of the denser CGM described here. However, until the full development of the SKA, FIR tracers of H{\sc i} gas such as the [C {\sc ii}] line will remain the only way to probe such phase at $z>1$. Promising avenues to detect the CGM at $z>3$ will be through [C{\sc ii}] line intensity mapping surveys, to be carried out, e.g., with CCAT-prime (\citep{Stacey+18}, see also Kovetz et al. A2020 WP), TIME \citep{Crites+17}, or CONCERTO \citep{Lagache+18}. These experiments will detect [C {\sc ii}] both from the host and from the diffuse CGM, without resolving individual structures, hence requiring higher-resolution follow ups. The CGM is also one of the main science goals of Athena (2030s), which will trace the warm/hot phase, of the Simons Observatory \citep{SimonsObs} and CMB-S4 \citep{CMB_S4} (2020s), which will utilize the thermal and kinematic SZ effects to probe ionized gas, and of LUVOIR (2030s), which will detect the CGM in absorption in the UV band.

\newpage

\bibliographystyle{unsrturltrunc6}
\bibliography{references}

\end{document}